\documentclass[twocolumn,floatfix,aps,prb]{revtex4}
\usepackage{graphicx}
\usepackage{amsmath}
\usepackage{amssymb}
\usepackage{bm}
\usepackage{hyperref}

\usepackage{color}

\begin{document}

\title{Low--high voltage duality in tunneling spectroscopy of the Sachdev-Ye-Kitaev model}
\author{N. V. Gnezdilov}
\affiliation{Instituut-Lorentz, Universiteit Leiden, P.O. Box 9506, 2300 RA Leiden, The Netherlands}
\author{J. A. Hutasoit}
\affiliation{Instituut-Lorentz, Universiteit Leiden, P.O. Box 9506, 2300 RA Leiden, The Netherlands}
\author{C. W. J. Beenakker}
\affiliation{Instituut-Lorentz, Universiteit Leiden, P.O. Box 9506, 2300 RA Leiden, The Netherlands}
\date{July 2018}

\begin{abstract}
The Sachdev-Ye-Kitaev (SYK) model describes a strongly correlated metal with all-to-all random interactions (average strength $J$) between $N$ fermions (complex Dirac fermions or real Majorana fermions). In the large-$N$ limit a conformal symmetry emerges that renders the model exactly soluble. Here we study how the non-Fermi liquid behavior of the closed system in equilibrium manifests itself in an open system out of equilibrium. We calculate the current-voltage characteristic of a quantum dot, described by the complex-valued SYK model, coupled to a voltage source via a single-channel metallic lead (coupling strength $\Gamma$). A one-parameter scaling law appears in the large-$N$ conformal regime, where the differential conductance $G=dI/dV$ depends on the applied voltage only through the dimensionless combination $\xi=eVJ/\Gamma^2$. Low and high voltages are related by the duality $G(\xi)=G(\pi/\xi)$. This provides for an unambiguous signature of the conformal symmetry in tunneling spectroscopy.
\end{abstract}
\maketitle

\textit{Introduction ---} The Sachdev-Ye-Kitaev model, a fermionic version\cite{Kitaev} of a disordered quantum Heisenberg magnet,\cite{Sachdev, Sachdev3} describes how $N$ fermionic zero-energy modes are broadened into a band of width $J$ by random infinite-range interactions. The phase diagram of the SYK Hamiltonian can be solved exactly in the large-$N$ limit,\cite{Sachdev2,Maldacena,Polchinski} when a conformal symmetry emerges at low energies that forms a holographic description of the horizon of an extremal black hole in a (1+1)-dimensional anti-de Sitter space.\cite{Kitaev,Sachdev3,Sachdev2,Kit18}

To be able to probe this holographic behaviour in the laboratory, it is of interest to create a ``black hole on a chip",\cite{Pikulin,Alicea,Pikulin2} that is, to realize the SYK model in the solid state. Ref.\ \onlinecite{Pikulin} proposed to use a quantum dot formed by an opening in a superconducting sheet on the surface of a topological insulator. In a perpendicular magnetic field the quantum dot can trap vortices, each of which contains a Majorana zero-mode.\cite{Fu08} Chiral symmetry ensures that the band only broadens as a result of four-Majorana-fermion terms in the Hamiltonian, a prerequisite for the real-valued SYK model. A similar construction uses an array of Majorana nanowires coupled to a quantum dot.\cite{Alicea} Since it might be easier to start from conventional electrons rather than Majorana fermions, Ref.\ \onlinecite{Pikulin2} suggested to work with the complex-valued SYK model of interacting Dirac fermions in the zeroth Landau level of a graphene quantum dot. Chiral symmetry at the charge-neutrality point again suppresses broadening of the band by two-fermion terms.

\begin{figure}[tb!]
\centerline{\includegraphics[width=0.9\linewidth]{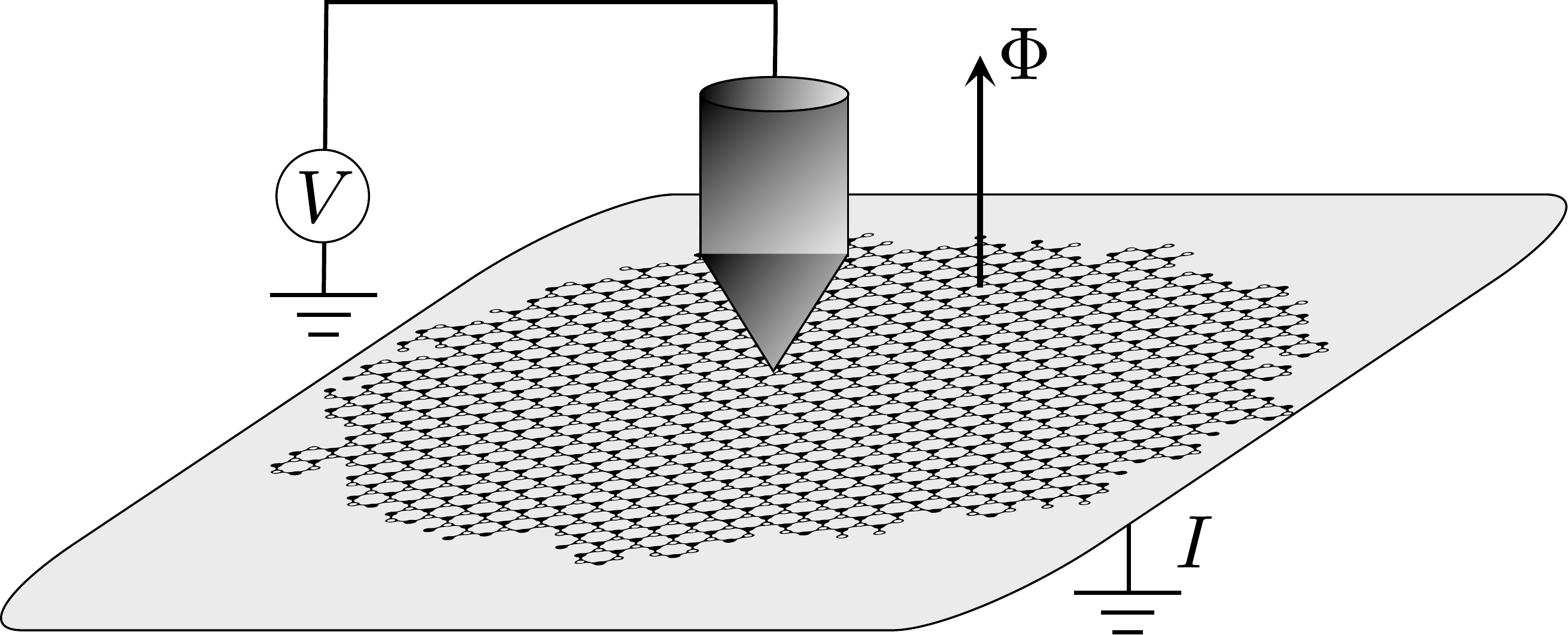}}
\caption{Tunneling spectroscopy of a graphene flake, in order to probe the complex-valued SYK model.\cite{Pikulin2} We calculate the current $I$ driven by a voltage $V$ through a single-channel point contact (coupling strength $\Gamma$) into a graphene flake on a grounded conducting substrate. At the charge neutrality point a chiral symmetry ensures that the zeroth Landau level (degeneracy $N=e\Phi/h$ for an enclosed flux $\Phi$) is only broadened by electron-electron interactions (strength $J$). For a sufficiently random boundary the quantum dot can be described by the SYK Hamiltonian \eqref{H}.
}
\label{fig_layout}
\end{figure}

The natural way to study a quantum dot is via transport properties. Electrical conduction through chains of SYK quantum dots has been studied in Refs.\ \onlinecite{Balents,Sachdev_chain,Zha17,Gu17,Che17,Ben18,Hal18,Zho18}. For a single quantum dot coupled to a tunnel contact, as in Fig.\ \ref{fig_layout}, Refs.\ \onlinecite{Pikulin,Alicea,Pikulin2} studied the limit of negligibly small coupling strength $\Gamma$, in which the differential conductance $G=dI/dV$ equals the density of states of the quantum dot. Conformal symmetry in the large-$N$ limit gives a low-voltage divergence $\propto 1/\sqrt{V}$, until $eV$ drops below the single-particle level spacing $\delta\simeq J/N$.\cite{Bagrets, Bagrets2,Feigelman}

Here we investigate how a finite $\Gamma$ affects the tunneling spectroscopy. We focus on the complex-valued SYK model for Dirac fermions, as in the graphene quantum dot of Ref.\ \onlinecite{Pikulin2}. Our key result is that in the large-$N$ conformal symmetry regime $J/N\ll eV\ll J$ the zero-temperature differential conductance of the quantum dot depends on $\Gamma, J$, and $V$ only via the dimensionless combination $\xi=eVJ/\Gamma^2$. Low and high voltages are related by the duality $G(\xi)=G(\pi/\xi)$, providing an experimental signature of the conformal symmetry. 
 
\textit{Tunneling Hamiltonian ---}
We describe the geometry of Fig.\ \ref{fig_layout} by the Hamiltonian 
\begin{align}  
&{}{H}= {}{H}_{\rm SYK} +\sum_p \varepsilon_p {}{\psi}^\dag_p    {}{\psi}^{\vphantom{\dagger}}_p +
\sum_{i,p}( \lambda_i {}{c}_i^\dag  {}{\psi}^{\vphantom{\dagger}}_p +\lambda_i^\ast {}{\psi}_p^\dag  {}{c}^{\vphantom{\dagger}}_i),\nonumber\\
&{H}_{\rm SYK}= (2 N)^{-3/2}{\textstyle\sum_{ijkl}} J_{ij;kl} \, {}{c}^\dag_i {}{c}^\dag_j {}{c}^{\vphantom{\dagger}}_k {}{c}^{\vphantom{\dagger}}_l,\label{H}\\
&J_{ij;kl}=J^*_{kl;ij}=-J_{ji;kl}=-J_{ij;lk}.\nonumber
\end{align}
The annihilation operators $c_i$, $i=1,2,\ldots$ represent the $N=h\Phi/e$ interacting Dirac fermions in the spin-polarized zeroth Landau level of the graphene quantum dot (enclosing a flux $\Phi$). Two-fermion terms $c_{i}^\dagger c_{j}$ are suppressed by chiral symmetry when the Fermi level $\mu=0$ is at the charge-neutrality point (Dirac point).\cite{Pikulin2} The operators $\psi_p$ represent electrons at momentum $p$ in the single-channel lead (dispersion $\varepsilon_p=p^2/2m$, linearized near the Fermi level), coupled to mode $i$ in the quantum dot with complex amplitude $\lambda_i$. The tunneling current depends only on the sum of $|\lambda_i|^2$, via the coupling strength
\begin{equation}
\Gamma=\pi\rho_{\rm lead} \,{\textstyle\sum_{i}}|\lambda_{i}|^2,\;\;\rho_{\rm lead}=(2\pi\hbar v_{\rm F})^{-1}.\label{Gammadef}
\end{equation}
If ${\cal T}\in(0,1)$ is the transmission probability into the quantum dot, one has $\Gamma\simeq {\cal T}N\delta\simeq {\cal T}J$. 

The Hamiltonian $H_{\rm SYK}$ is the complex-valued SYK model \cite{Sachdev2} if we take random couplings $J_{ij;kl}$ that are independently distributed Gaussians with zero mean $\langle{J_{ij;kl}}\rangle=0$ and variance $\langle{|J_{ij;kl}|^2}\rangle=J^2$. The zeroth Landau level then broadens into a band of width $J$, corresponding to a single-particle level spacing $\delta\simeq J/N$ (more precisely, $\delta\simeq J/N\ln N$).\cite{Bagrets} In the energy range $\delta\ll \varepsilon\ll J$ the retarded Green's functions can be evaluated in saddle-point approximation,\cite{Sachdev2}
\begin{equation}
G^{\rm R}(\varepsilon)= -\mathrm{i} \pi^{1/4}\sqrt{\frac{\beta}{2 \pi J}}\,  \frac{\bm{\Gamma}\left(1/4-\mathrm{i}\beta \varepsilon/2 \pi  \right)}{\bm{\Gamma}\left(3/4-\mathrm{i}\beta \varepsilon/2 \pi \right)} \label{GR}, 
\end{equation}
where $\beta=1/k_{\rm B}T$ and $\bm{\Gamma}(x)$ is the Gamma function. At zero temperature this simplifies to
\begin{equation}
G^{\rm R}(\varepsilon)=-\mathrm{i}  \pi^{1/4} \, \exp\left[ \tfrac{1}{4} \mathrm{i}\pi\,\mathrm{sgn}(\varepsilon) \right]\left|J \varepsilon\right|^{-1/2}. \label{GR_T0}
\end{equation}
Quantum fluctuations around the saddle point cut off the low-$\varepsilon$ divergence for $|\varepsilon|<\delta$.\cite{Bagrets, Bagrets2,Feigelman}

\textit{Tunneling current ---} The quantum dot is strongly coupled to a grounded substrate,\cite{Altman} so the current is entirely determined by the transmission of electrons through the point contact. The current operator $\bm{I}$ is given by the commutator
\begin{equation}
{}{\bm{I}} = \frac{\mathrm{i} e}{\hbar} \left[{}{H}, {\textstyle\sum_{p}} {}{\psi}^\dag_p {}{\psi}_p \right]=  \mathrm{i}\frac{e}{\hbar}{\sum_{n,p}} \left( \lambda_n {}{c}_n^\dag {}{\psi}_p -\lambda_n^\ast {}{\psi}_p^\dag {}{c}_n \right).\label{I_def}
\end{equation}
We calculate the time-averaged expectation value of $\bm{I}$ using the Keldysh path integral technique,\cite{Keldysh,Levitov,Kamenev,Kamenev2} which has previously been applied to the SYK model in Refs.\ \onlinecite{Balents, Zha17,Hal18,Sachdev5}.
The expectation value $I$ of the tunneling current is given by the first derivative of cumulant generating function:\cite{Levitov}
\begin{align}
{}&I= -\mathrm{i} \lim_{\chi\rightarrow 0} \frac{\partial}{\partial \chi} \ln Z(\chi) \, \label{I_av}, \\
{}&Z(\chi)= \bigl\langle \mathcal{T}_C \exp\bigl( - \mathrm{i} \textstyle{\int_C} \mathrm{d} t \left[ {}{H} +\tfrac{1}{2}\chi(t)\bm{I} \right] \bigr) \bigr\rangle  \label{Z}.
\end{align}
Here $\mathcal{T}_C$ indicates time-ordering along the Keldysh contour\cite{Keldysh} of the counting field $\chi(t)$, equal to $+\chi$ on the forward branch of the contour (from $t=0$ to $t=\infty$) and equal to $-\chi$ on the backward branch (from $t=\infty$ to $t=0$). 
The calculation is worked out in the Appendix. 

\begin{figure}[tb!]
\centerline{\includegraphics[width=1\linewidth]{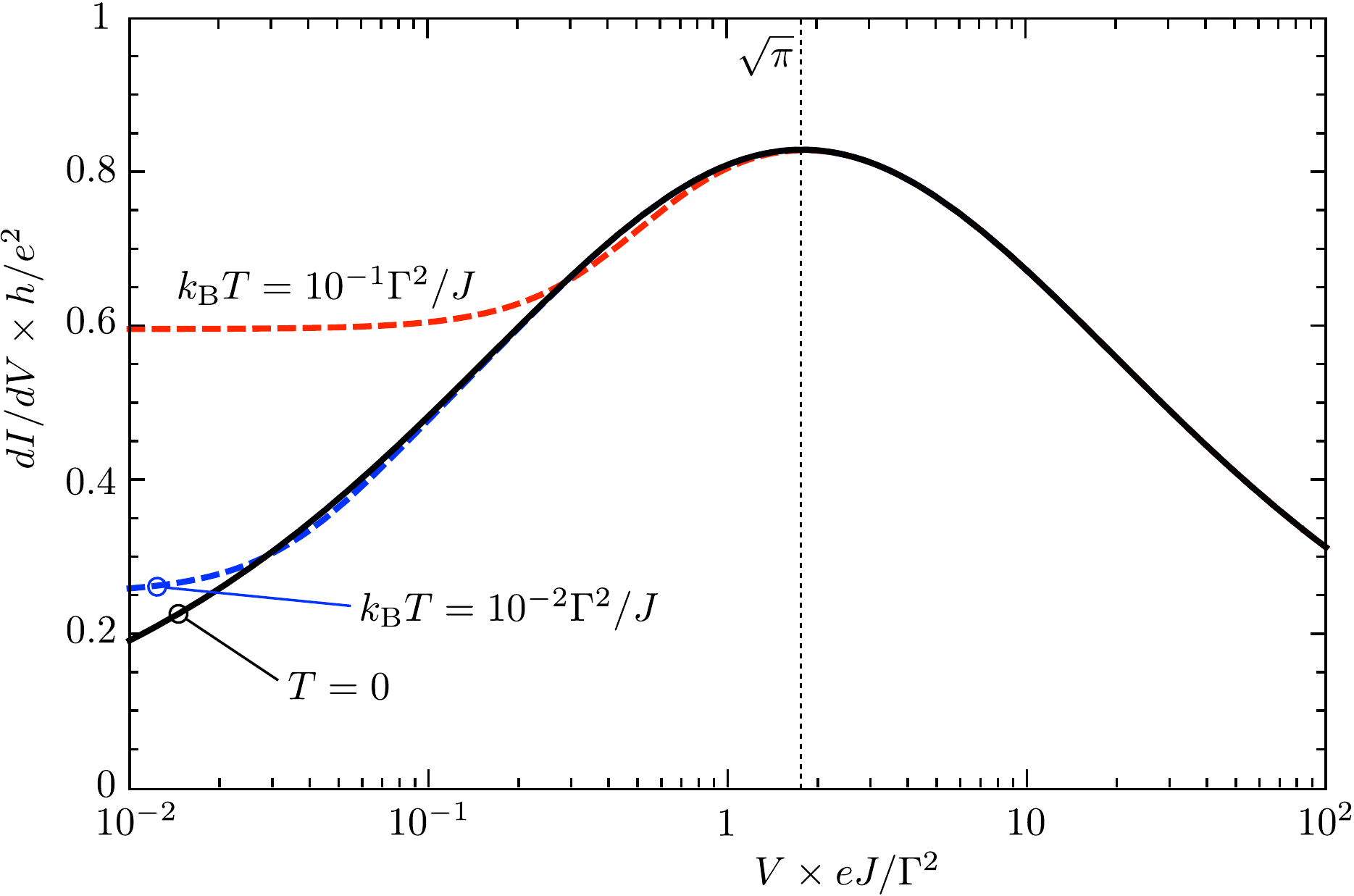}}
\caption{Differential conductance $G=dI/dV$ calculated from Eq.\ \eqref{dIdV_T}, as a function of dimensionless voltage $\xi=eVJ/\Gamma^2$ for three different temperatures. On the semi-logarithmic scale the duality between low and high voltages shows up as a reflection symmetry along the dotted line (where $\xi=\sqrt\pi$).
}
\label{fig_dIdV}
\end{figure}

The result for the differential conductance is
\begin{align}
G=\frac{d I}{d V}=\frac{e^2}{h}\int_{-\infty}^{+\infty} \mathrm{d} \varepsilon \, f'(\varepsilon-eV) \frac{4  \Gamma \,{\rm Im}\,G^{\rm R}(\varepsilon)}{|1+\mathrm{i} \Gamma  G^{\rm R}(\varepsilon)|^2}\, \label{dIdV_T},
\end{align}
where $f(\varepsilon)=(1+e^{\beta\varepsilon})^{-1}$ is the Fermi function. Substitution of the  conformal Green's function \eqref{GR} gives upon integration the finite temperature curves plotted in Fig.\ \ref{fig_dIdV}. 

At zero temperature $f'(\varepsilon-eV)\rightarrow -\delta(\varepsilon-eV)$ and substitution of Eq.\ \eqref{GR_T0} produces a single-parameter function of $\xi=eVJ/\Gamma^2$,
\begin{equation}
G(\xi)=\frac{e^2}{h}\frac{2\sqrt{2}}{\sqrt{2}+\pi^{1/4}\xi^{-1/2}+\pi^{-1/4}\xi^{1/2}}.\label{Gxiresult}
\end{equation}

\textit{Low--high voltage duality ---} The $T=0$ differential conductance \eqref{Gxiresult} in the conformal regime $J/N\ll eV\ll J$ satisfies the duality relation
\begin{equation}
G(\xi)=G(\pi/\xi),\;\;\text{if}\;\;N^{-1}(J/\Gamma)^2\ll\xi,1/\xi\ll(J/\Gamma)^2.\label{Gduality}
\end{equation}
The $V$-to-$1/V$ duality is visible in the semi-logarithmic Fig.\ \ref{fig_dIdV} by a reflection symmetry of the differential conductance along the $\xi=\sqrt\pi$ axis. The symmetry is precise at $T=0$, and is broken in the tails with increasing temperature.

The voltage range in which $V$ and $1/V$ are related by Eq.\ \eqref{Gduality} covers the full conformal regime for $N\simeq (J/\Gamma)^4$. In this voltage range the $1/\sqrt V$ tail at high voltages crosses over to a $\sqrt V$ decay at low voltages. The high-voltage tail reproduces the $1/\sqrt V$ differential conductance that follows\cite{Pikulin,Alicea,Pikulin2} from the density of states in the limit $\Gamma\rightarrow 0$ (since $\xi\rightarrow\infty$ for $\Gamma\rightarrow 0$). The density of states gives\cite{Bagrets, Bagrets2,Feigelman} a crossover to a $\sqrt V$ decay when $eV$ drops below the single-particle level spacing $\delta\simeq J/N$. Our finite-$\Gamma$ result \eqref{Gxiresult} implies that this crossover already sets in at larger voltages $eV\simeq\Gamma^2/J$, well above $\delta$ for $N\gg(J/\Gamma)^2$.

The symmetrically peaked profile of Fig.\ \ref{fig_dIdV} is a signature of conformal symmetry in as much as this produces a power-law singularity in the retarded propagator at low energies. It is not specific for the square-root singularity \eqref{GR_T0}, other exponents would give a qualitatively similar low-high voltage duality. For example, the generalized SYK$_{2p}$ model with $2p\geq 4$ interacting Majorana fermion terms has a $\varepsilon^{(1-p)/p}$ singularity,\cite{Maldacena, Sachdev_chain} corresponding to the duality $G(\xi_p)=G(C_p/\xi_p)$ with $C_p$ a numerical coefficient and $\xi_p=(eV)^{2(p-1)/p} J^{2/p}  \Gamma^{-2}$. In contrast, a disordered Fermi liquid such as the non-interacting SYK$_2$ model, with Hamiltonian $H=\sum_{ij} J_{ij} c_i^\dag c_j$, has a constant propagator at low energies and hence a constant $dI/dV$ in the range $J/N \ll eV \ll J$.

\textit{Conclusion ---} We have shown that tunneling spectroscopy can reveal a low--high voltage duality in the conformal regime of the Sachdev-Ye-Kitaev model of $N$ interacting Dirac fermions. A physical system in which one might search for this duality is the graphene quantum dot in the lowest Landau level, proposed by Chen, Ilan, De Juan, Pikulin, and Franz.\cite{Pikulin2} 

As argued by those authors, one should be able to reach $N$ of order $10^2$ for laboratory magnetic field strengths in a sub-micrometer-size quantum dot. This leaves two decades in the conformal regime $J/N\ll eV\ll J$. If we tune the tunnel coupling strength near the ballistic limit $\Gamma\lesssim J$, it should be possible even for these moderately large values of $N$ to achieve $N\simeq (J/\Gamma)^4$ and access the duality over two decades of voltage variation. For such large $\Gamma$ the condition on temperature, $k_{\rm B}T\ll\Gamma^2/J$ would then also be within experimental reach ($J\simeq 34\,{\rm meV}$ from Ref.\ \onlinecite{Pikulin2} and $\Gamma\simeq 10\,{\rm meV}$ has $k_{\rm B}T=10^{-2}\,\Gamma^2/J$ at $T=300\,{\rm mK}$).

\textit{Acknowledgements ---}
We have benefited from discussions with K. E. Schalm and A. Romero Bermudez. This research was supported by the Netherlands Organization for Scientific Research (NWO/OCW)  and by an ERC Synergy Grant.

\appendix

\section{Outline of the calculation}

We describe the calculation leading to Eq.\ \eqref{dIdV_T} for the current-voltage characteristics, generalizing it to nonzero chemical potential $\mu$ and including also the shot noise power. We set $\hbar$ and $e$ to unity, except for the final formulas.

\subsection{Generating function of counting statistics}

Arbitrary cumulants of the current operator \eqref{I_def} can be obtained from the generating function \eqref{Z}.
A gauge transformation allows us to write equivalently
\begin{align}
&Z(\chi)= \bigl\langle \mathcal{T}_C \exp\bigl( - \mathrm{i} \textstyle{\int_C}   {}{H}(t)\,\mathrm{d} t\bigr) \bigr\rangle,\\
&{}{H}(t)= {}{H}_{\rm SYK}+\textstyle{\sum_p} \varepsilon_p {}{\psi}^\dag_p    {}{\psi}^{\vphantom{\dagger}}_p-\mu\textstyle{\sum_n}  c_n^\dagger c_n^{\vphantom{\dagger}}\nonumber\\
& +
\textstyle{\sum_{n,p}}\left( \mathrm{e}^{\mathrm{i}  \chi(t)/2}\lambda_n {}{c}_n^\dag  {}{\psi}^{\vphantom{\dagger}}_p +\mathrm{e}^{-\mathrm{i}  \chi(t)/2}\lambda_n^\ast {}{\psi}^{{\dagger}}_p {}{c}_n^{\vphantom{\dagger}}\right).
\end{align}
For generality we have added a chemical potential term $\propto\mu$. (In the main text we take $\mu=0$, corresponding to a quantum dot at charge neutrality.)

We need the advanced and retarded Green's functions $G^{\rm A}(\varepsilon)=\bigl(G^{\rm R}(\varepsilon)\bigr)^*$ and the Keldysh Green's function
\begin{equation}
G^{\rm K}(\varepsilon)=\mathcal{F}(\varepsilon)\left(G^{\rm R}(\varepsilon)-G^{\rm A}(\varepsilon)\right),\;\;
\mathcal{F}(\varepsilon)=\tanh (\beta \varepsilon/2).
\end{equation}
These are collected in the matrix Green's function ${\cal G}$, which on the Keldysh contour has the representation\cite{Kamenev,Kamenev2,LO_APP}
\begin{align}
&{\cal G}=\begin{pmatrix}
G^{\rm R} & G^{\rm K} \\ 0 & G^{\rm A}
\end{pmatrix} = L \sigma_3 \begin{pmatrix}
G^{++} & G^{+-} \\ G^{-+} & G^{--}
\end{pmatrix} L^{\dag}, \label{A_G} \\
&L=\frac{1}{\sqrt{2}} \begin{pmatrix}
1 & -1 \\ 1 & 1
\end{pmatrix},\;\;\sigma_3=\begin{pmatrix}
1&0\\
0&-1
\end{pmatrix},
\end{align}
in terms of the Green's functions on the forward and backward branches of the contour:
\begin{subequations}
\label{GKeldysh}
\begin{align}
&{}G^{++}(t, t')=-\mathrm{i}N^{-1}\sum_n \left\langle \mathcal{T}\,  {}{c}_n(t) {}{c}^{\dag}_n(t') \right\rangle,\\
&G^{+-}(t, t')=\mathrm{i}N^{-1}\sum_n \left\langle {}{c}^{\dag}_n(t')   {}{c}_n(t) \right\rangle ,\\
&{}G^{-+}(t, t')=-\mathrm{i}N^{-1}\sum_n \left\langle  {}{c}_n(t) {}{c}^{\dag}_n(t')  \right\rangle, \\
&G^{--}(t, t')=-\mathrm{i}N^{-1}\sum_n \left\langle \mathcal{T}^{-1}\,  {}{c}_n(t) {}{c}^{\dag}_n(t') \right\rangle. \label{A_G_def}
\end{align}
\end{subequations}
The operators $\mathcal{T}$ and $\mathcal{T}^{-1}$ order the times in increasing and decreasing order, respectively. 

\subsection{Saddle point solution}

In the regime $J/N\ll\varepsilon\ll J$ the Green's function of the SYK model is given by the saddle point solution\cite{Sachdev2}
\begin{align}
G^{\rm R}(\varepsilon)= -\mathrm{i} C\mathrm{e}^{-\mathrm{i}\theta}\sqrt{\frac{\beta}{2 \pi J}}\,  \frac{\Gamma\left(\tfrac{1}{4}-\mathrm{i}\tfrac{\beta \varepsilon}{2 \pi} + \mathrm{i}\mathcal{E} \right)}{\Gamma\left(\tfrac{3}{4}-\mathrm{i}\tfrac{\beta \varepsilon}{2 \pi} + \mathrm{i}\mathcal{E}\right)},\label{GRapp}
\end{align}
with the definitions
\begin{equation}
\mathrm{e}^{2 \pi \mathcal{E}}=\frac{\sin\left(\tfrac{\pi}{4}+\theta\right)}{\sin\left(\tfrac{\pi}{4}-\theta\right)},\;\;
C=\left(\pi/\cos 2 \theta\right)^{1/4}.
\end{equation}
The angle $\theta\in(-\pi/4,\pi/4)$ is a spectral asymmetry angle,\cite{Parcollet_APP} determined by the charge per site ${\cal Q}\in(-1/2,1/2)$ on the quantum dot according to\cite{Sachdev4_APP}
\begin{equation}
\mathcal{Q}=N^{-1}\textstyle{\sum_i }\langle {}{c}_i^\dag {}{c}_i \rangle-\tfrac{1}{2}=-{\theta}/{\pi}-\tfrac{1}{4}{\sin 2 \theta}.
\end{equation}
For $\mu=0$, when ${\cal Q}=0$, one has $\theta=0$, $C=\pi^{1/4}$. In good approximation (accurate within 15\%), 
\begin{equation}
\theta\approx - \tfrac{1}{2}\pi {\cal Q}\Rightarrow C\approx(\pi/\cos\pi {\cal Q})^{1/4}.\label{thetaQrelation}
\end{equation}

In the mean-field approach the quartic SYK interaction \eqref{H} is replaced by a quadratic one with the kernel ${\cal G}^{-1}$ from Eq.\ \eqref{A_G}. A Gaussian integration over the Grassmann fields gives the generating function
\begin{widetext}
\begin{equation} 
\ln Z= \int_{-\infty}^{\infty}\frac{\mathrm{d} \varepsilon}{2 \pi}\ln\left(\frac{\det\left[1-\Gamma\Xi(\varepsilon)  \Lambda^\dag {\cal G}(\varepsilon) \Lambda\right]}{\det\left[1-\Gamma \Xi(\varepsilon) {\cal G}(\varepsilon)\right]}\right),  \;\;\Lambda=  \begin{pmatrix}
\cos \left( \chi/2\right) & \mathrm{i} \sin \left( \chi/2\right) \\ \mathrm{i} \sin \left( \chi/2\right) & \cos \left( \chi/2\right)
\end{pmatrix},\;\;\Xi(\varepsilon)=-\mathrm{i}  \begin{pmatrix}
1 &  2  \mathcal{F}(\varepsilon-V)  \\ 0 & -1
\end{pmatrix}.
\end{equation}
The matrix $\Xi(\varepsilon)$ is the Keldysh Green's function of the lead, integrated over the momenta. This evaluates further to
\begin{equation}
 \ln Z= \int_{-\infty}^{\infty}\frac{d \varepsilon}{2 \pi}\ln\left[ 1 +\frac{\mathrm{i} \Gamma\left(G^{\rm R}-G^{\rm A}\right)}{\left(1+\mathrm{i} \Gamma G^{\rm R}\right)\left(1-\mathrm{i} \Gamma G^{\rm A}\right)}  \bigg(\left[1- \mathcal{F}(\varepsilon) \mathcal{F}(\varepsilon-V)\right]\left(\cos \chi -1 \right) +\mathrm{i} \left[\mathcal{F}(\varepsilon)-\mathcal{F}(\varepsilon-V) \right] \sin \chi \bigg) \right] .
\end{equation}
\end{widetext}
At zero temperature the distribution function simplifies to
$\mathcal{F}(\varepsilon) \mapsto \mathrm{sgn}\left(\varepsilon\right)$, hence
\begin{equation} 
\ln Z =  \int_{0}^{V}\frac{d \varepsilon}{2 \pi}\ln\left[ 1 +\frac{2 \mathrm{i} \Gamma \left(G^{\rm R}-G^{\rm A}\right)\left(\mathrm{e}^{\mathrm{i} \chi} -1 \right)}{\left(1+\mathrm{i} \Gamma G^{\rm R}\right)\left(1-\mathrm{i} \Gamma G^{\rm A}\right)}  \right].  \label{CGFT0}
\end{equation}

\subsection{Average current and shot noise power}

A $p$-fold differentation of $Z(\chi)$ with respect to $\chi$ gives the $p$-th cumulant of the current. In this way the full counting statistics of the charge transmitted through the quantum dot can be calculated.\cite{Levitov} The first cumulant, the time-averaged current $I$ from Eq.\ \eqref{I_av}, is given by
\begin{equation} 
I{}=\frac{e}{h}\int_{-\infty}^{+\infty} d \varepsilon \,  \frac{\mathrm{i}  \Gamma \left[\mathcal{F}(\varepsilon)-\mathcal{F}(\varepsilon-V) \right] \left(G^{\rm R}-G^{\rm A}\right)}{\left(1+\mathrm{i} \Gamma  G^{\rm R}\right)\left(1-\mathrm{i} \Gamma  G^{\rm A}\right)},
\end{equation}
which is Eq.\ \eqref{dIdV_T} from the main text.

At zero temperature the differential conductance $G=dI/dV$ is
\begin{align}
G(\xi) = \frac{2 e^2}{h}\left[1+\frac{1}{ 2\sin\left(\pi/4+\theta\right)}\left(\frac{\sqrt{\xi}}{C}+\frac{C}{\sqrt{\xi}}\right)\right]^{-1},
\end{align}
with $\xi=eVJ/\Gamma^2$. The duality relation
\begin{equation}
G(\xi)=G(C^4/\xi)
\end{equation}
reduces to the one from the main text, $G(\xi)=G(\pi/\xi)$, when we set $\mu=0\Rightarrow \theta=0\Rightarrow C=\pi^{1/4}$.

The second cumulant, the shot noise power $P$, follows similarly from
\begin{equation}
P=  -\lim_{\chi\rightarrow 0} \frac{\partial^2}{\partial \chi^2} \ln Z(\chi).
\end{equation} 
The Fano factor $F$, being the ratio of the shot noise power and the current at zero temperature, is simply given by
\begin{equation} 
F=\frac{dP/dV}{dI/dV}=e\left(1 -\frac{h}{e^2} G\right) .
\end{equation}
It has the same one-parameter scaling and duality as $G$. The fact that higher order cumulants of the current have the same scaling as the differential conductance is a consequence of the single-point-contact geometry, with a single counting field $\chi(t)$. This does not carry over to a two-point-contact geometry.


\begin{thebibliography}{99}
\bibitem{Kitaev} A. Kitaev, {\it A simple model of quantum holography}, 
KITP Program: Entanglement in Strongly-Correlated Quantum Matter, \href{http://online.kitp.ucsb.edu/online/entangled15/kitaev/}{April
7} and  \href{http://online.kitp.ucsb.edu/online/entangled15/kitaev2/}{May 27}, 2015.
\bibitem{Sachdev} S. Sachdev and J. Ye, {\it Gapless spin-fluid ground state in a random quantum Heisenberg magnet}, \href{https://doi.org/10.1103/PhysRevLett.70.3339}{Phys. Rev. Lett. {\bf 70}, 3339 (1993)}.
\bibitem{Sachdev3} S. Sachdev, {\it Holographic metals and the fractionalized Fermi liquid}, \href{https://doi.org/10.1103/PhysRevLett.105.151602}{Phys. Rev. Lett. {\bf 105}, 151602 (2010)}.
\bibitem{Sachdev2} S. Sachdev, {\it Bekenstein-Hawking entropy and strange metals}, \href{https://doi.org/10.1103/PhysRevX.5.041025}{Phys. Rev. X {\bf 5}, 041025 (2015)}.
\bibitem{Maldacena} 
J. Maldacena and D. Stanford, {\it Remarks on the Sachdev-Ye-Kitaev model}, \href{https://doi.org/10.1103/PhysRevD.94.106002}{Phys. Rev. D {\bf 94}, 106002 (2016)}.
\bibitem{Polchinski}
J. Polchinski and V. Rosenhaus, {\it The spectrum in the Sachdev-Ye-Kitaev model}, \href{https://doi.org/10.1007/JHEP04(2016)001}{JHEP \textbf{04}, 1 (2016)}.
\bibitem{Kit18} A. Kitaev and S. J. Suh, \textit{The soft mode in the Sachdev-Ye-Kitaev model and its gravity dual}, \href{https://doi.org/10.1007/JHEP05(2018)183}{JHEP \textbf{05}, 183 (2018)}.
\bibitem{Pikulin} D. I. Pikulin and M. Franz, {\it Black hole on a chip: proposal for a physical realization of the SYK model in a solid-state system}, \href{https://doi.org/10.1103/PhysRevX.7.031006}{Phys. Rev. X {\bf 7}, 031006 (2017)}.
\bibitem{Alicea} A. Chew, A. Essin, and J. Alicea, {\it Approximating the Sachdev-Ye-Kitaev model with Majorana wires}, \href{https://doi.org/10.1103/PhysRevB.96.121119}{Phys. Rev. B {\bf 96}, 121119(R) (2017)}.
\bibitem{Pikulin2} A. Chen, R. Ilan, F. de Juan, D. I. Pikulin, and M. Franz, {\it  Quantum holography in a graphene flake with an irregular boundary}, \href{https://doi.org/10.1103/PhysRevLett.121.036403}{Phys. Rev. Lett. \textbf{121}, 036403 (2018)}.
\bibitem{Fu08} L. Fu and C. L. Kane, \textit{Superconducting proximity effect and Majorana fermions at the surface of a topological insulator}, \href{https://doi.org/10.1103/PhysRevLett.100.096407}{Phys. Rev. Lett. \textbf{100}, 096407 (2008)}.
\bibitem{Balents} X.-Y. Song, C.-M. Jian, and L. Balents, {\it Strongly correlated metal built from Sachdev-Ye-Kitaev models}, \href{https://doi.org/10.1103/PhysRevLett.119.216601}{Phys. Rev. Lett. {\bf 119}, 216601 (2017)}.
\bibitem{Sachdev_chain} R. A. Davison, W. Fu, A. Georges, Y. Gu, K. Jensen, and S. Sachdev, {\it Thermoelectric transport in disordered metals without quasiparticles: The Sachdev-Ye-Kitaev models and holography}, \href{https://doi.org/10.1103/PhysRevB.95.155131}{Phys. Rev. B {\bf 95}, 155131 (2017)}.
\bibitem{Zha17} P. Zhang, \textit{Dispersive SYK model: band structure and quantum chaos}, 
\href{https://doi.org/10.1103/PhysRevB.96.205138}{Phys. Rev. B {\bf 96}, 205138 (2017)}.
\bibitem{Gu17} Y. Gu, X.-L. Qi, and D. Stanford, \textit{Local criticality, diffusion and chaos in generalized Sachdev-Ye-Kitaev models}, \href{https://doi.org/10.1007/JHEP05(2017)125}{JHEP \textbf{05}, 125 (2017)}.
\bibitem{Che17} Xin Chen, Ruihua Fan, Yiming Chen, Hui Zhai, and Pengfei Zhang, \textit{Competition between chaotic and non-chaotic phases in a quadratically coupled Sachdev-Ye-Kitaev model}, \href{https://doi.org/10.1103/PhysRevLett.119.207603}{Phys. Rev. Lett. \textbf{119}, 207603 (2017)}.
\bibitem{Ben18} D. Ben-Zion and J. McGreevy, \textit{Strange metal from local quantum chaos}, \href{https://doi.org/10.1103/PhysRevB.97.155117}{Phys. Rev. B {\bf 97}, 155117 (2018)}.
\bibitem{Hal18} A. Haldar, S. Banerjee, and V. B. Shenoy, \textit{Higher-dimensional SYK non-Fermi liquids at Lifshitz transitions}, \href{https:doi.org/10.1103/PhysRevB.97.241106}{Phys. Rev. B \textbf{97}, 241106 (2018)}.
\bibitem{Zho18} Y. Zhong, \textit{Periodic Anderson model meets Sachdev-Ye-Kitaev interaction: A solvable playground for heavy fermion physics}, \href{https://arxiv.org/abs/1803.09417}{arXiv:1803.09417}
\bibitem{Bagrets}
D. Bagrets, A. Altland, and A. Kamenev, {\it Sachdev-Ye-Kitaev model as Liouville quantum mechanics}, \href{https://doi.org/10.1016/j.nuclphysb.2016.08.002}{Nucl. Phys. B {\bf 911}, 191 (2016)}.
\bibitem{Bagrets2} 
D. Bagrets, A. Altland, A. Kamenev,  {\it Power-law out of time order correlation functions in the SYK model}, \href{https://doi.org/10.1016/j.nuclphysb.2017.06.012}{Nucl. Phys. B {\bf 921}, 727 (2017)}.
\bibitem{Feigelman} A. V. Lunkin, K. S. Tikhonov, and M. V. Feigel'man, {\it SYK model with quadratic perturbations: the route to a non-Fermi-liquid}, \href{https://arxiv.org/abs/1806.11211}{arXiv:1806.11211}.
\bibitem{Altman} We assume that the grounded substrate does not spoil the non-Fermi liquid state of the quantum dot. This might happen if the coupling becomes too strong, according to S. Banerjee and E. Altman, {\it Solvable model for a dynamical quantum phase transition from fast to slow scrambling}, \href{https://doi.org/10.1103/PhysRevB.95.134302}{Phys. Rev. B {\bf 95}, 134302 (2017)}.
\bibitem{Keldysh} L. V. Keldysh, {\it Diagram technique for nonequilibrium processes}, \href{http://www.jetp.ac.ru/cgi-bin/e/index/e/20/4/p1018?a=list}{JETP {\bf 20}, 4, p. 1018 (1965)}.
\bibitem{Levitov}  L. S. Levitov, H.-W. Lee, and G. B. Lesovik, {\it Electron counting statistics and coherent states of electric current}, \href{https://doi.org/10.1063/1.531672}{Journal of Mathematical Physics {\bf 37}, 4845 (1996)}.
\bibitem{Kamenev} 
A. Kamenev and A. Levchenko, {\it Keldysh technique and non-linear $\sigma$-model: basic principles and applications}, \href{https://doi.org/10.1080/00018730902850504}{Adv. in Physics {\bf 58}, 197 (2009)}.
\bibitem{Kamenev2} A. Kamenev, {\it Field Theory of Non-Equilibrium Systems} (Cambridge University Press, 2011).
\bibitem{Sachdev5} A. Eberlein, V. Kasper, S. Sachdev, and J. Steinberg, {\it Quantum quench of the Sachdev-Ye-Kitaev Model}, \href{https://doi.org/10.1103/PhysRevB.96.205123}{Phys. Rev. B {\bf 96}, 205123 (2017)}.
\bibitem{LO_APP} A. I. Larkin and Yu. N. Ovchinnikov, {\it Nonlinear conductivity of superconductors in the mixed state}, \href{http://www.jetp.ac.ru/cgi-bin/e/index/e/41/5/p960?a=list}{JETP {\bf 41}, 960 (1975)}.
\bibitem{Parcollet_APP} O. Parcollet, A. Georges, G. Kotliar, and A. Sengupta,
{\it Overscreened multichannel SU(N) Kondo model: Large-N solution and conformal field theory}, \href{https://doi.org/10.1103/PhysRevB.58.3794}{Phys. Rev. B {\bf 58}, 3794 (1998)}.
\bibitem{Sachdev4_APP} A. Georges, O. Parcollet, and S. Sachdev, {\it Quantum fluctuations
of a nearly critical Heisenberg spin glass}, \href{https://doi.org/10.1103/PhysRevB.63.134406}{Phys. Rev. B {\bf 63}, 134406 (2001)}.

\end{thebibliography}
\end{document}